\documentclass[%
 reprint,
showpacs,
 amsmath,amssymb,
 aps,
 pra,
longbibliography,
floatfix
]{revtex4-1}

\usepackage[utf8]{inputenc}	
\usepackage[T1]{fontenc}	
\usepackage{graphicx}		
\usepackage{color}
\usepackage[english]{babel}
\usepackage{hyperref}
\usepackage{url}
\usepackage{todonotes}
\usepackage{lipsum}
\usepackage{bm}

\usepackage{braket}

\usepackage[position=top,caption=false]{subfig}

\let\originaleqref\eqref
\renewcommand{\eqref}{Eq.~\originaleqref}

\newcommand{\fref}[1]{\figurename~\ref{#1}}

\renewcommand{\vec}[1]{\bm{#1}}

\newcommand{\e}[1]{\mathrm{e}^{#1}}

\newcommand{\Tr}[1]{\mathrm{Tr}\left(#1\right)}

\newcommand{\pdiff}[2]{\frac{\partial #1}{\partial #2}}

\renewcommand{\L}[0]{\mathcal{L}}

\newcommand{\rst}[0]{\rho_\mathrm{ss}}

\newcommand{\tp}{t^\prime}
\renewcommand{\L}{\mathcal{L}}

\newcommand{\dT}{\delta_{\text{PDS}}}
\newcommand{\dTh}{\delta_{T}}
\newcommand{\dL}{\delta_{\text{Q}}}
\newcommand{\dM}{\delta_{\text{max}}}
\newcommand{\PT}{P_{\text{trap}}}

\newcommand{\dLor}{\delta_{\text{L}}}

\graphicspath{
{./Figures/}
}

\begin{document}
\title{Random search for a dark resonance}
\author{Alexander Holm Kiilerich}
\email{kiilerich@phys.au.dk}
\author{Klaus Mølmer}
\email{moelmer@phys.au.dk}
\date{\today}
\affiliation{Department of Physics and Astronomy, Aarhus University, Ny Munkegade 120, DK 8000 Aarhus C. Denmark}
\date{\today}

\bigskip

\begin{abstract}
A pair of resonant laser fields can drive a three-level system into a dark state where it ceases to absorb and emit radiation due to destructive interference.
We propose a scheme to search for this resonance by randomly changing the frequency of one of the fields each time a fluorescence photon is detected. The longer the system is probed, the more likely the frequency is close to resonance and the system populates the dark state. Due to the correspondingly long waiting times between detection events, the evolution is non-ergodic and the precision of the frequency estimate does not follow from the conventional Cram\'{e}r-Rao bound of parameter estimation. Instead, a L\'{e}vy statistical analysis yields  the scaling of the estimation error with time for precision probing of this kind.
\end{abstract}

\maketitle
\noindent

\section{Introduction}
Quantum systems can act as sensitive probes and field sensors \cite{PhysRevLett.96.010401}, and since measurements yield random outcomes, the precision by which the value of a physical parameter can be determined follows from a statistical analysis.
For $N$ repeated, independent measurements, the estimation error is governed by the Cramér Rao bound \cite{Cramer} and the Fisher information \cite{Fisher}, and scales as it $1/\sqrt{N}$.

Recent works have addressed the complementary situation of continuous measurements on a single quantum system, and it was recognized that, e.g., photon counting in a fluorescence experiment of duration $T$ is equivalent to $N \propto T$ independent measurements of the waiting time between consecutive detector clicks \cite{PhysRevA.89.052110,PhysRevA.91.012119}.
The back action of continuously performed measurements on a quantum system triggers  transient evolution witnessed in the signal correlation functions  \cite{PhysRevA.94.032103} and if they have finite relaxation time, the estimation error based on the signal mean values and two-time correlations scales as $1/\sqrt{T}$ \cite{PhysRevA.94.032103,Burgarth2015}.

\begin{figure}
\subfloat[\label{fig:setup}]{
\includegraphics[trim=0 0 0 0,width=0.95\columnwidth]{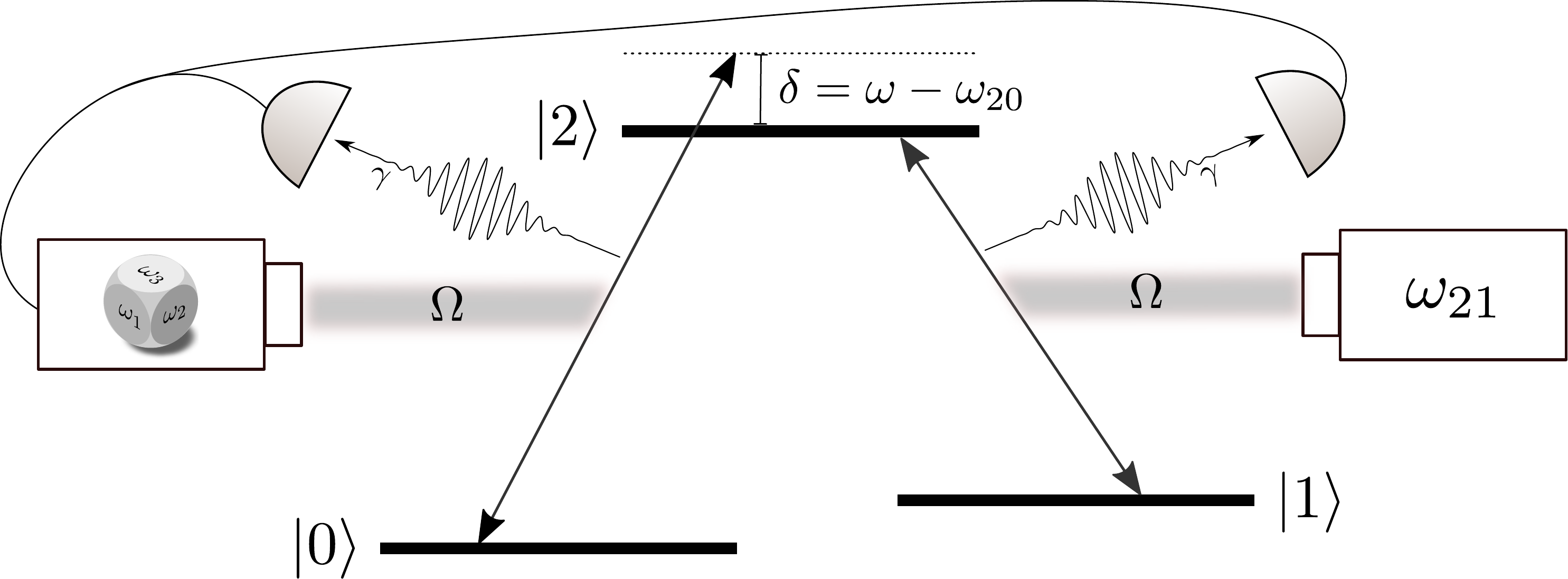}
}

\subfloat[\label{fig:trajectory}]{
\includegraphics[trim=0 0 0 0,width=0.95\columnwidth]{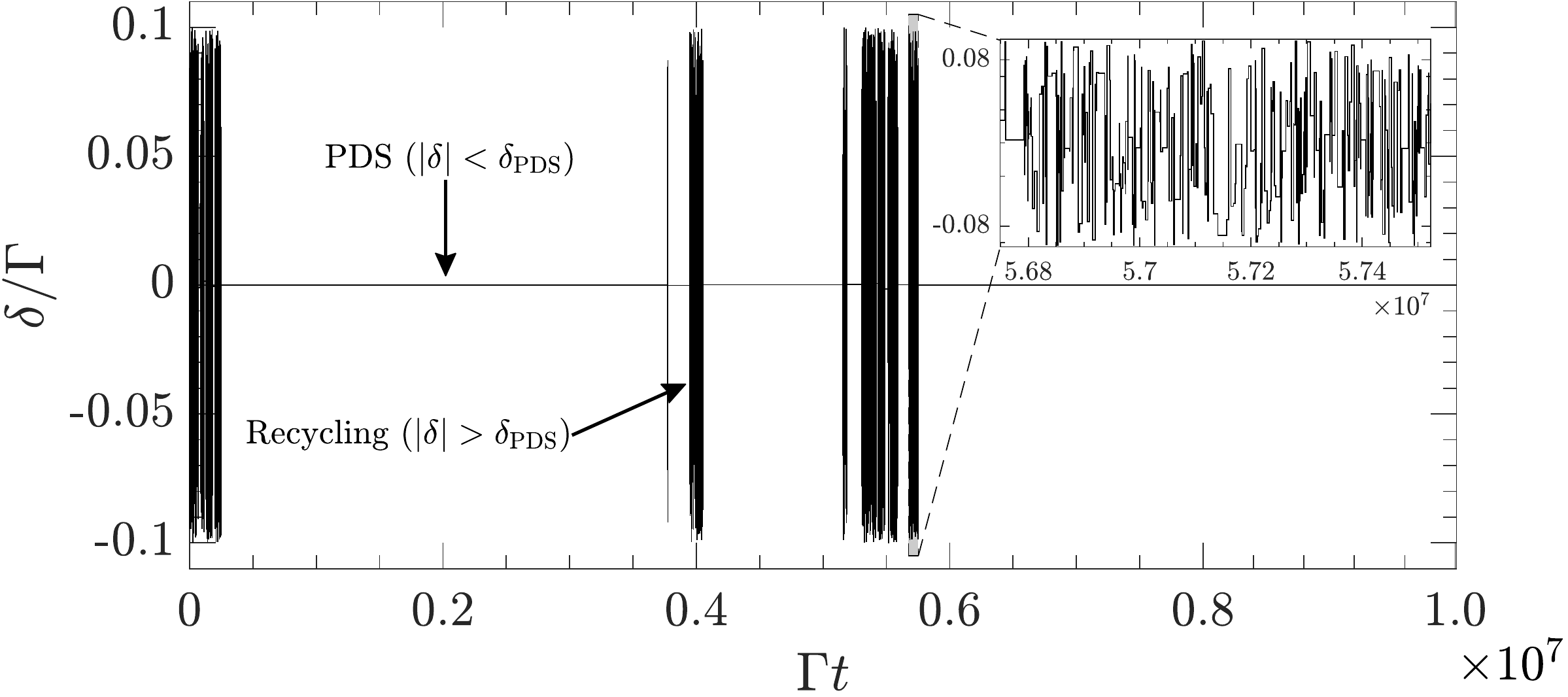}
}
\caption{(a) $\Lambda$-type system driven by laser fields with Rabi frequency $\Omega$. The $\ket{1}\leftrightarrow\ket{2}$ coupling laser is kept on resonance while the $\ket{0}\leftrightarrow\ket{1}$ coupling laser is detuned by an amount $\delta= \omega-\omega_{20}$, where $\omega_{20}$ is the atomic resonance frequency. Both emission channels are monitored by photo detectors, and upon detection in either channel $\delta$ is shifted randomly on a uniform interval with $\delta\in [-\dM,\dM]$.
(b) Quantum Monte Carlo simulated trajectory for the detuning $\delta$ as a function of time $t$. The simulation is made with $\Omega=0.1\Gamma/\sqrt{2}$ and $\dM=0.1\Gamma$ where $\Gamma^{-1}$ is the excited state lifetime.
}

\label{fig:Figure1}
\end{figure}

In this paper, we consider the special case where the fluorescence rate of an atomic system vanishes when it is excited by a laser field on exact resonance. Such dark resonances occur in connection with the phenomenon of electromagnetically induced transparency \cite{Gray:78,PhysRevLett.66.2593}, and due to their narrow linewidths, they are sensitive probes of perturbations on the system; see, e.g., \cite{0295-5075-44-1-031,6910267,0295-5075-54-3-323}.  As an alternative to a systematic scanning and accumulation of signal at different, discrete laser frequencies, we investigate a random search protocol in which the probe laser frequency may come arbitrarily close to the dark resonance. That event is witnessed by the complete absence of signal and suggests application of the following adaptive protocol for the duration $T$ of the experiment:
The system is excited at a frequency picked uniformly within a fixed interval, including the resonance.
Whenever a photon is detected, a new random laser frequency is chosen and the system is excited until the next photodetection, where the frequency is again shifted.
The protocol is illustrated for a driven $\Lambda$-type system in \fref{fig:setup}.

When driven far from the dark resonance, the high scattering rate implies a high probability for an early photon detection and a shift to a different frequency, while for frequencies close to resonance, the photon emission rate is very small, and these frequencies are hence maintained for a long time before the next emission event. We thus expect that the longer we probe the atom, the more likely are occurrences of long intervals with laser frequencies close to the dark resonance. The instantaneous, stochastically tuned
laser frequency thus constitutes a good estimate of the atomic transition frequency. Due to the distribution of short, long, and very long time intervals, however, the dynamics is not ergodic,  and the Cram\'{e}r-Rao bound which relies on asymptotic normality can neither be used to assess the quantitative achievements of the protocol nor to estimate how the error scales with the duration of the experiment.

We show here that the problem is tractable by methods of generalized statistics \cite{PaulLevy,bouchaud1990anomalous} that have been developed
to analyze non-ergodic dynamics in, e.g.,
animal foraging behavior \cite{sims2008scaling,viswanathan1999optimizing}, human travel patterns \cite{brockmann2006scaling}, earthquake occurrences \cite{PhysRevLett.97.178501} and financial systems \cite{10.2307/2350970,mantegna1996turbulence}.
In quantum physics  they have found applications in analysis of
anomalous transport properties of quantum arrays \cite{PhysRevB.72.075309}, and our approach is inspired by and closely follows Bardou \textit{et. al.} \cite{PhysRevLett.72.203}, who apply L\'{e}vy statistics to subrecoil laser cooling mediated by a dark state mechanism. While we provide quantitative results and simulations for a specific model, the analysis is general, and we shall return to wider consequences and applications of our results in the final sections of the paper.

In Section \ref{sec:model}, we introduce the atomic model and illustrate our random search protocol by performing a quantum trajectory analysis of the photon counting and random frequency shifts. In Section \ref{sec:LevyAnalysis}, we present a Lévy statistical analysis of the search protocol.
We give criteria for the success of our protocol as an estimation strategy and analyze the scaling of the estimation precision with time. In Section \ref{sec:Comparison}, we compare the random search protocol to a systematic scan across a dark resonance.
Finally, in Section\ref{sec:Outlook}, we provide an outlook on the generality of our derivations and the applicability of our results to similar schemes and other systems with dark resonances.
The Appendix includes background material and derivations of the central results in from main text.

\section{Atomic model and trajectory analysis}\label{sec:model}
\begin{figure}
\includegraphics[trim=0 0 0 0,width=0.95\columnwidth]{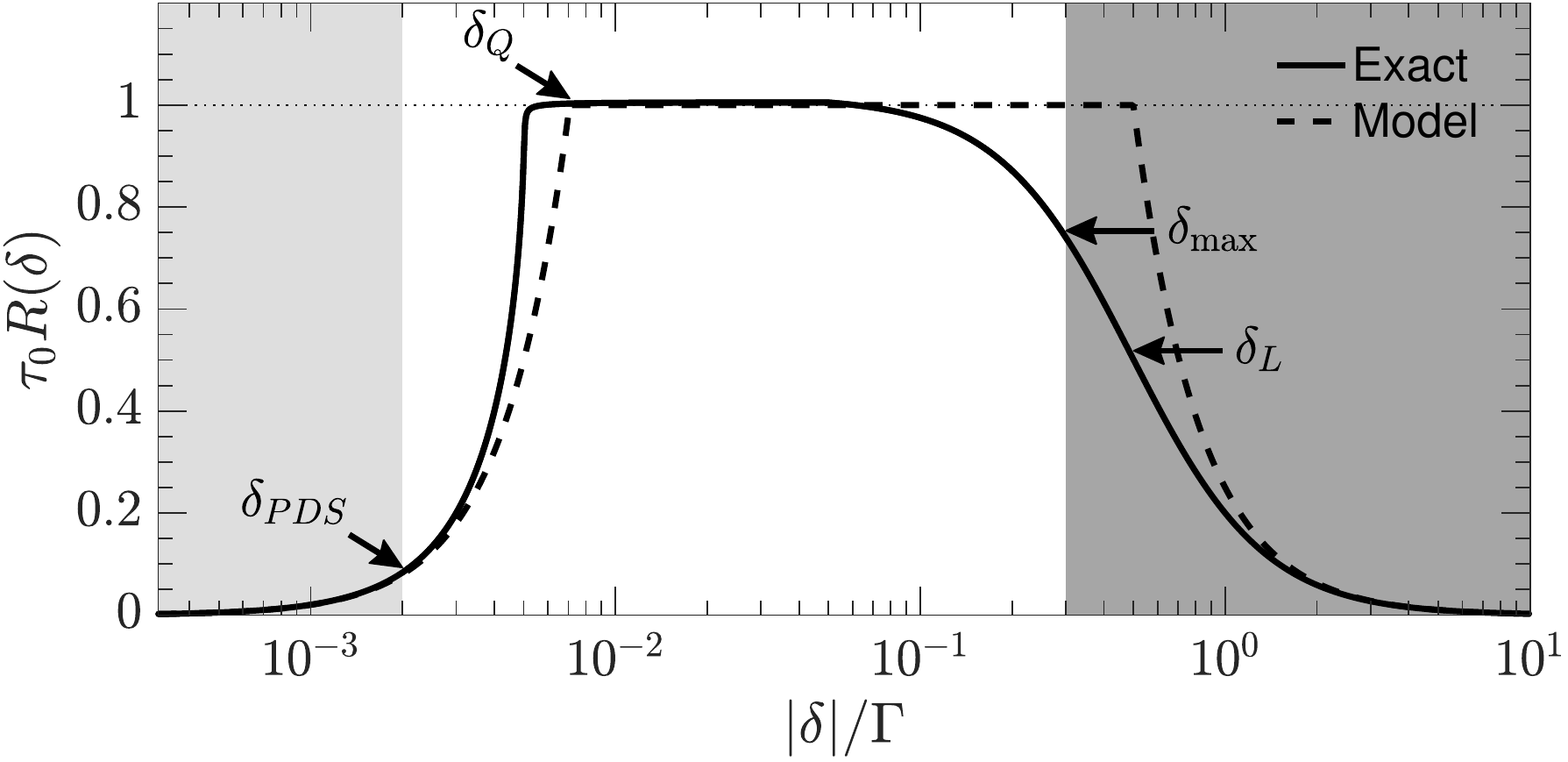}
\caption{Effective frequency dependent photo emission rate from the dark state $\ket{\psi_-}$ shown for $\Omega/\Gamma = 0.1\Gamma/\sqrt{2}$. The full line shows the exact rate and dashed the line our simplified model \eqref{eq:rate}. Characteristic detunings (see main text) are annotated. The rate is an even function of $\delta$. The light shaded area is the trapping region and the dark shaded area marks the frequency range not included in the stochastic scan.
}
\label{fig:rate}
\end{figure}

\fref{fig:setup} depicts the situation of a $\Lambda$-type three-level quantum system interacting with two laser fields with equal coupling strengths. Assume that one field is fixed on resonance, while the other is scanned with a detuning $\delta=\omega-\omega_{20}$ from the exact resonance in the system. The upper level is unstable and decays with equal probabilities into the two low-lying states, which can both be expanded on the dark state $\ket{\psi_-} = (|0\rangle-|1\rangle)/\sqrt{2}$ and the bright state $\ket{\psi_+}=(|0\rangle+|1\rangle)/\sqrt{2}$. The bright state is coupled to the excited state, and after a short time, the system starting in state $|0\rangle$ or $|1\rangle$ has either undergone excitation and emitted a photon or been effectively projected into the dark state \cite{PhysRevLett.68.580}. The dark state has a vanishing excitation rate but for a finite detuning, the phase difference between the laser and the dark state atomic components evolves, and leads to an effective photon emission rate $R(\delta)$. This rate is derived in Appendix \ref{sec:R} and shown as a function of the detuning $\delta$ in \fref{fig:rate}.
If the coupling laser is tuned slightly away from resonance, the effective photo emission rate depends quadratically on the detuning $\delta$, and for a range  $|\delta| < \delta_{\mathrm{PDS}}$, the system will be trapped for a long time in a pseudo-dark state (PDS). At higher detunings the excitation rate levels off and decreases when the detuning exceeds $\delta_L \simeq \Gamma$, the excited state linewidth.

A characteristic waiting time between subsequent emissions is $\tau(\delta) = 1/R(\delta)$.
Ergodicity relies on the ability to average single trajectories over long times
compared to any intrinsic time scale, but since $R(\delta)\rightarrow 0$ we have $\tau(\delta)\rightarrow \infty$ for $\delta\rightarrow 0$,
so even a very long time $T$ may be dominated by a single waiting time with $|\delta|<\dTh$, where $R(\dTh)T=1$.

We shall restrict the choice of frequencies to an interval $|\delta| < \dM$, containing the resonance, but avoiding the wings of the absorption profile, $\delta_{\mathrm{max}} < \delta_L$.
To verify the intuition behind the scheme, we show in \fref{fig:trajectory} the evolution of the detuning as a function of time as obtained from a Monte Carlo wave function simulation of the continuous measurements and random frequency jumps \cite{PhysRevLett.68.580}. The total duration $T$ is, indeed, dominated by a few long intervals with small detuning, interrupted by brief periods with larger, fluctuating values of $\delta$. The value of the laser frequency at any random time is likely to be very close to the atomic resonance frequency.

To obtain analytic predictions for the generic behavior of our estimation protocol, we shall focus in the following section on the most significant features and abandon less important details.
The variation of the fluorescence rate $R(\delta)$ by an atom occupying the pseudo dark state $\ket{\psi_-}$ will thus be approximated by the function
 \begin{equation} \label{eq:rate}
 R(\delta)=\left\{
             \begin{array}{ll}
               \tau_0^{-1}(\delta/\dL)^2, &\quad |\delta| < \dL \\
               \tau_0^{-1}, &\quad \dL < |\delta| < \dLor \\
               \tau_0^{-1}(\dLor/\delta)^2,  &\quad \dLor < |\delta|.
             \end{array}
           \right.
\end{equation}
This simple form of $R(\delta)$, illustrated by the dashed curve in \fref{fig:rate}, is adequate to represent the very long and very short waiting times attained for $\delta\simeq 0$ and for larger $\delta$, respectively. The parameters, yielding the best agreement with the actual rate for the $\Lambda$-system illustrated by the solid curve in \fref{fig:rate}, are derived in Appendix \ref{sec:R}.

\section{Lévy statistical analysis}\label{sec:LevyAnalysis}
While the simulation illustrates the apparent success of such an estimation strategy, a quantitative analysis of its precision and its scaling with $T$ is hampered by the fact that the probability distribution $P(\tau)$ of dwell times $\tau$ between detection events has a very long tail, and its mean and variance formally diverge in the interesting regime where $\delta\rightarrow 0$. For such problems, e.g., the sum of $N$ waiting times $T_N=\sum_{i=1}^N \tau^{(i)}$ does not obey the central limit theorem (CLT) and will not converge to a Gaussian distributed variable with a mean value proportional to $N$. Instead, the increasing probability that a single term attains a very large value and dominates the sum may cause it to scale as a higher power of $N$. This is the characteristic property of L\'{e}vy flights, and $P(T_N)$ is a Lévy distribution \cite{bouchaud1990anomalous}.

In \fref{fig:trajectory}, we see how the evolution is comprised of two different time scales:
In a narrow interval $|\delta|\leq  \dT \leq  \dL$,  the system occupies the PDS for which the waiting times are of the order $\tau\propto \delta^{-2}$.
A single detector click here will with overwhelming probability cause a jump to a detuning $|\delta|\geq \dT$ where the waiting times are short and many jumps occur before the system returns to the narrow PDS detuning interval.
A trajectory thus consists of a number of trapping intervals $\tau_t^{(1)},\tau_t^{(2)},\dots$ interspersed by recycling periods of duration $\tau_r^{(1)},\tau_r^{(2)},\dots$ each containing many detection events. The competition between trapping and recycling periods is at the core of our statistical analysis, and the probability distributions $P_t(\tau_t)$ of trapping times $\tau_t$ and $P_r(\tau_r)$  of recycling intervals $\tau_r$ will suffice to analyze the asymptotic behavior of our estimation scheme as $T\rightarrow\infty$.

For $P_t(\tau_t)$ we note that since each detuning in the PDS interval is reached with equal probability, the density of trajectories just returned to the PDS is $\rho(\delta)=\frac{1}{2}\dT^{-1}$. Upon return with a given $\delta$, the probability of a trapping time $\tau_t$ is ascribed by the delay function $w(\tau_t|\delta)$ which can be calculated by a master equation analysis \cite{CARMICHAEL,PhysRevA.89.052110}. In the limit of predominantly long waiting times, $w(\tau_t|\delta)$ is well approximated by a single exponential function, $w(\tau_t|\delta) = R(\delta)^{-1}\exp(-\tau_t R(\delta))$, where the frequency-dependent emission rate $R(\delta)$ vanishes at $\delta=0$; cf. \eqref{eq:rate}.

The distribution $P_t(\tau_t)$ of trapping intervals is given by integrating $w(\tau_t|\delta)$ over the PDS region with weight $\rho(\delta)$, and for long $\tau_t$ we find
\begin{equation} \label{eq:power}
P_t(\tau_t) \underset{\text{large } \tau_t}{\simeq} \frac{\mu \tau_b^{\mu}}{\tau_t^{1+\mu}},
\end{equation}
where $\mu=1/2$, and $\tau_b=\tau_0\pi\left(\dL/\dT\right)^2/16$.
As anticipated by the arguments above, $P_t(\tau_t)$ decreases very slowly (as $1/\tau_t^{3/2}$) for large values of $\tau$, and we are in the regime where standard Gaussian statistics must be replaced by L\'{e}vy statistics.

For a distribution with power-law tails such as \eqref{eq:power}, all moments $\braket{\tau^n}$ for which $n\geq \mu$ diverge.
A well-known example is a Cauchy distribution, which has $\mu=1$.
The central limit theorem of Gaussian statistics states that
for $\mu>1$ the total time spent in the trapping region $T_{N}^{(\text{PDS})} = \sum_{i=1}^N \tau_t^{(i)}$ is proportional to $N$, while for $\mu<1$ any sequence is dominated by rare events and the generalized CLT dictates that asymptotically $T_N^{(\mathrm{PDS})}\propto N^{1/\mu}$. See Appendix \ref{sec:ABroad} for a brief introduction to broad distributions and the generalized CLT.

The behavior of $R(\delta)$ for large $\delta$ determines $P_r(\tau_r)$.
When setting up the protocol, we have a choice in the maximum and minimum values allowed in the random selection of $\delta$ after each detection event.
We assume that a rough prior estimate restricts the search interval $\delta\in[-\dM,\dM]$ around $\omega=\omega_{20}$. The symmetry is not of importance since we assume $\dM\gg \dT$.
The properties of $P_r(\tau_r)$ depend on the value of $\dM$ compared to the characteristic detunings $\dL$ and $\dLor$.
If $\dL \ll \dM<\dLor$, the high $\delta$ rate is given by the plateau in \fref{fig:rate}, and as derived in Appendix \ref{sec:Arecycling} we obtain a finite mean value $\braket{\tau_r}=\tau_0(\dM/\dT)$, implying that $T_N^{(\mathrm{REC})}= \sum_{i=1}^N \tau_r^{(i)}$ grows linearly with $N$. For simplicity, we restrict our attention to this case and defer discussion of the case with $\dM>\dLor$ to Section \ref{sec:Outlook}.

\subsection{Trapped proportion}
The results for the trapping times and recycling intervals already provide qualitative insight regarding the asymptotic achievements of our estimation scheme at large times $T$ (large $N$). For $\dM<\dLor$, $T_N^{(\text{PDS})}\propto N^2$ dominates over $T_N^{(REC)}\propto N$, and we expect trajectories to spend most of the time occupying the PDS.
In fact, the time averaged proportion of time in the PDS is given by $f_{T}(T)=T_N^{(\text{PDS})}/(T_N^{(\text{PDS})}+T_N^{(\text{REC})})$, which by applying the generalized CLT (see Appendix \ref{sec:ABroad}) for long times $T$ can be written
$f_{T}(T) = 1-\xi(\braket{\tau_r}/\tau_b)T^{(\mu-1)}$.
This reveals a time-averaged non-PDS proportion decreasing as $1/\sqrt{T}$, but contrary to ergodic processes with Gaussian statistics it continues to fluctuate, via the Lévy increment $\xi$, even in the high-$T$ limit.

\begin{figure}
\includegraphics[trim=0 0 0 0,width=0.95  \columnwidth]{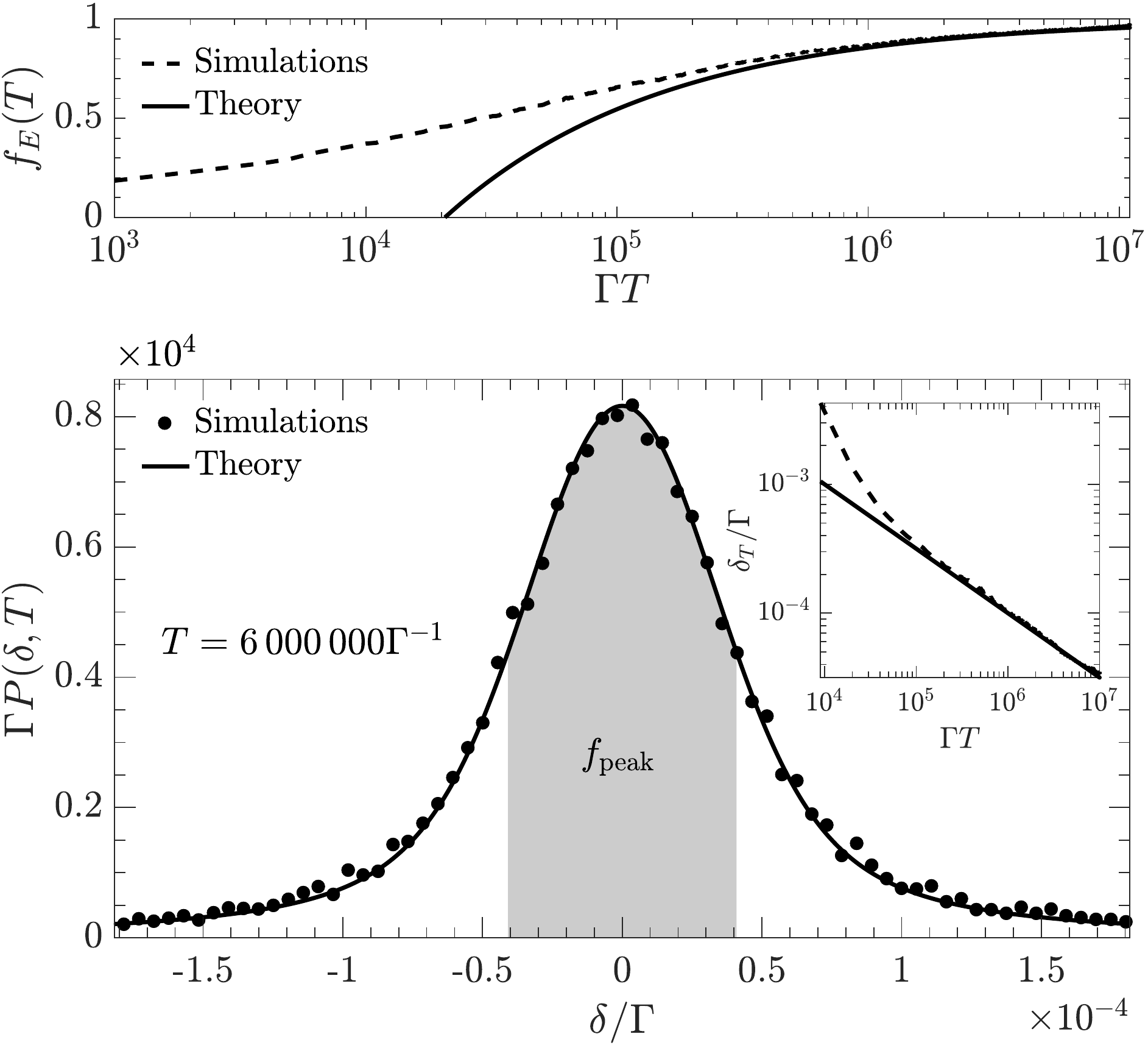}
\caption{
Top: Proportion of trapped trajectories \eqref{eq:f} (with $\dT=0.01\Gamma$) as a function of time. The dashed line depicts a quantum jump simulation of $20\,000$ trajectories with the 	same parameters as in \fref{fig:trajectory}. It matches the statistical model (full line) for (very) large times.
Bottom: Distribution of the detuning $\delta$ after a long time $T=6\times 10^6\Gamma^{-1}$. The dots show simulated data, the full line the theoretical result of our statistical analysis
and the shaded area marks the fraction with $|\delta|\leq \dTh$.
The inset shows how the characteristic width $\delta_T$ of the distribution scales as $T^{-1/2}$ and matches the model for times larger than $\sim 10^5\Gamma^{-1}$.
}
\label{fig:distribution}
\end{figure}

The ensemble averaged proportion of trajectories that will asymptotically be trapped in the PDS is derived in Appendix \ref{sec:AtrapProportion},
\begin{align}\label{eq:f}
f_E^{}(T) \simeq 1- \frac{\sin(\pi\mu)}{\pi}\frac{\braket{\tau_r}}{ \tau_b^\mu T^{1-\mu}},
\end{align}
where we see the same scaling with time $T$, but without fluctuations.
Equation (\ref{eq:f}) expresses the probability as a function of time that the laser frequency is within $\dT$ of the true resonance frequency, while with a probability $1-f_E(T)$ the frequency resides, at the time $T$, in the recycling region, and it will not be a good estimator of the resonance frequency.
The convergence of $f_E(T)$ to unity for large $T$ hence signifies that the random search is a successful estimation scheme. In \fref{fig:distribution} we show how $f_E(T)$ matches the ensemble average of trajectories such as the one in \fref{fig:trajectory} for large times, $T\gtrsim 10^6\Gamma^{-1}$.

\subsection{Asymptotic frequency distribution and estimation sensitivity}
To address the sensitivity of the random search we consider the distribution $\mathcal{P}(\delta,T)$ of trajectories with $|\delta|<\dT$. The Lévy statistical analysis in Appendix \ref{sec:AFrequencyDistribution}, reveals that $\mathcal{P}(\delta,T)$ can be factorized as $\mathcal{P}(\delta,T)=h(T)G(q)$, where $h(T)$ is the time-dependent height of the distribution, and $G(q)$, where $q=\delta/\dTh$, is a form factor.
It is a signature of the broken ergodicity that $\mathcal{P}(\delta,T)$ depends explicitly on $T$ and does not approach a stationary form even for very long times.
We find
$
h(T)=\left(\tau_{PDS}/\tau_b\right)^\mu \sin(\pi\mu)/\left(\pi\mu \dTh\right),
$
where $\tau_{PDS} = 1/R(\dT)$.
A general expression for the form factor is given in Appendix \ref{sec:AFrequencyDistribution}. It depends only on the value of $\mu$, and for $\mu=1/2$ it may be expressed as $G(q)=D(q)/q$, where $D(q)$ is the Dawson function.
The tails of $G(q)$ are Lorentzian $\sim 1/2q^2$ and much wider than those of a Gaussian while its maximum is flat compared to a Lorentzian.

The important detuning scale is, as anticipated, given by $\dTh = \dL(\tau_0/T)^\mu$.
This implies that $h(T)\propto T^{\mu}$, and the full width at half-maximum (FWHM) of $G(q)$ is $q_w\dTh\propto T^{-\mu}$, where for $\mu=1/2$ we find numerically $q_w\simeq 2.13$. Since the distribution has long tails, we define the fraction $f_{\text{peak}}=\int_{-\dTh}^{\dTh} d\delta \, \mathcal{P}(\delta,T)$ of occurrences of final detunings in the characteristic range $|\delta| < \dTh$, as a measure for the parameter estimation sensitivity and we find $f_{\text{peak}}\simeq 0.59$ independently of $T$. This shows that asymptotically a constant part of the trajectories are within $\dTh \propto T^{-\mu}$ of the true resonance frequency. Note that the sensitivity does not depend on the values of $\dM$ and $\dT$ as long as $\dT\ll \dM<\dLor$.
For the $\Lambda$-system with $\mu=1/2$ we hence find a $1/\sqrt{T}$ scaling of the sensitivity in our estimation protocol.
We note that $59\%$ of the distribution within $\dTh$ corresponds to an $\simeq 0.82$ sigma confidence level if $\mathcal{P}(\delta,T)$ was a normal distribution.

In \fref{fig:distribution} we show how the ensemble obtained from simulations until $T = 6\times 10^6\Gamma^{-1}$ is well represented by $\mathcal{P}(\delta,T)$. The inset shows the consistency of the theoretical result for $\dTh$ with numerical results obtained directly from the sampled
$\mathcal{P}(\delta,T)$ as a function of time.

\section{Comparison to a systematic scan}\label{sec:Comparison}
We have shown that under certain restrictions our estimation scheme is successful, but it remains to be seen if it outperforms standard spectroscopy methods in the same settings.
A typical way to determine a resonance frequency is by observing florescence as the laser frequency is systematically scanned over the relevant frequencies with equal time at each point. The spectrum is reconstructed from the integrated fluorescence signal at each frequency. Such a scheme lends it self to a standard analysis relying on the Cram\'{e}r-Rao bound in a manner similar to \cite{PhysRevA.77.043606,PhysRevA.91.012119}.
In this section we perform such an analysis and compare the performance of a systematic scan to our stochastic protocol.

Assume first that a scan of total duration $T$ consists in observing the fluorescence for a time $t=T/N$ at each of a set of $N$ discrete, equally spaced frequencies $\{\delta_k\}_{k=1}^{N}$ on the search interval $[-\dM,\dM]$.
A data record $D=[n_1,n_2,\dots,n_N]^\mathrm{T}$ obtained in a time $T$ then contains the total photocount $n_k$ at each discrete frequency. These are independently sampled, and we assume that for large $T$ they are normally distributed with means $\overline{n}_k$ and variances $v_k$.
The full data record $D$ then samples a multivariate normal distribution
$P(D|\theta) = \mathcal{N}\left(\vec{\mu},\vec{\Sigma}\right)$ with mean value vector
$\vec{\mu}=[\overline{n}_1,\overline{n}_2,\dots,\overline{n}_N]^\mathrm{T}$
and a diagonal covariance matrix with elements $\Sigma_{kk} =v_k$.

The Fisher information for estimating a parameter $\theta$ from such a distribution is well-known,
\begin{align}\label{eq:FisherMultivariate}
\mathcal{I}(\theta) =
\pdiff{\vec{\mu}^\mathrm{T}}{\theta}
\vec{\Sigma}^{-1}
\pdiff{\vec{\mu}}{\theta}
+\frac{1}{2}\Tr{\vec{\Sigma}^{-1}\pdiff{\vec{\Sigma}}{\theta}\vec{\Sigma}^{-1}\pdiff{\vec{\Sigma}}{\theta}},
\end{align}
yielding in this case
\begin{align}\label{eq:discreteFisher}
\mathcal{I}(\theta) = \sum_k \frac{1}{v_k} \left(\pdiff{\overline{n}_k}{\theta}\right)^2+\frac{1}{2} \sum_k \left(\frac{1}{v_k} \pdiff{v_k}{\theta}\right)^2.
\end{align}
The mean and variance of the photocount at each discrete frequency follow from the master equation (\ref{eq:liovillian}). The mean fluorescence is $\overline{n}_k = \frac{T}{N}\tilde{R}(\delta)$. The photocount variance stems from temporal signal fluctuations, and it can be expressed as
\begin{align}\label{eq:vk}
v_k = \overline{n}_k+2T\sum_i\int_{0}^{\infty}d\tau\, \tilde{G}_i^{(2)}(\tau),
\end{align}
where the sum runs over the distinct emission channels, and $\tilde{G}_i^{(2)}(\tau)= G_i^{(2)}(\tau)-\Tr{\hat{c}_i^\dagger\hat{c}_i\rst}^2$.
The last term in \eqref{eq:vk} determines the deviation from Possonian counting statistics.
\begin{figure}
\subfloat[\label{fig:compareA}]{
\includegraphics[trim=0 0 0 0,width=0.9\columnwidth]{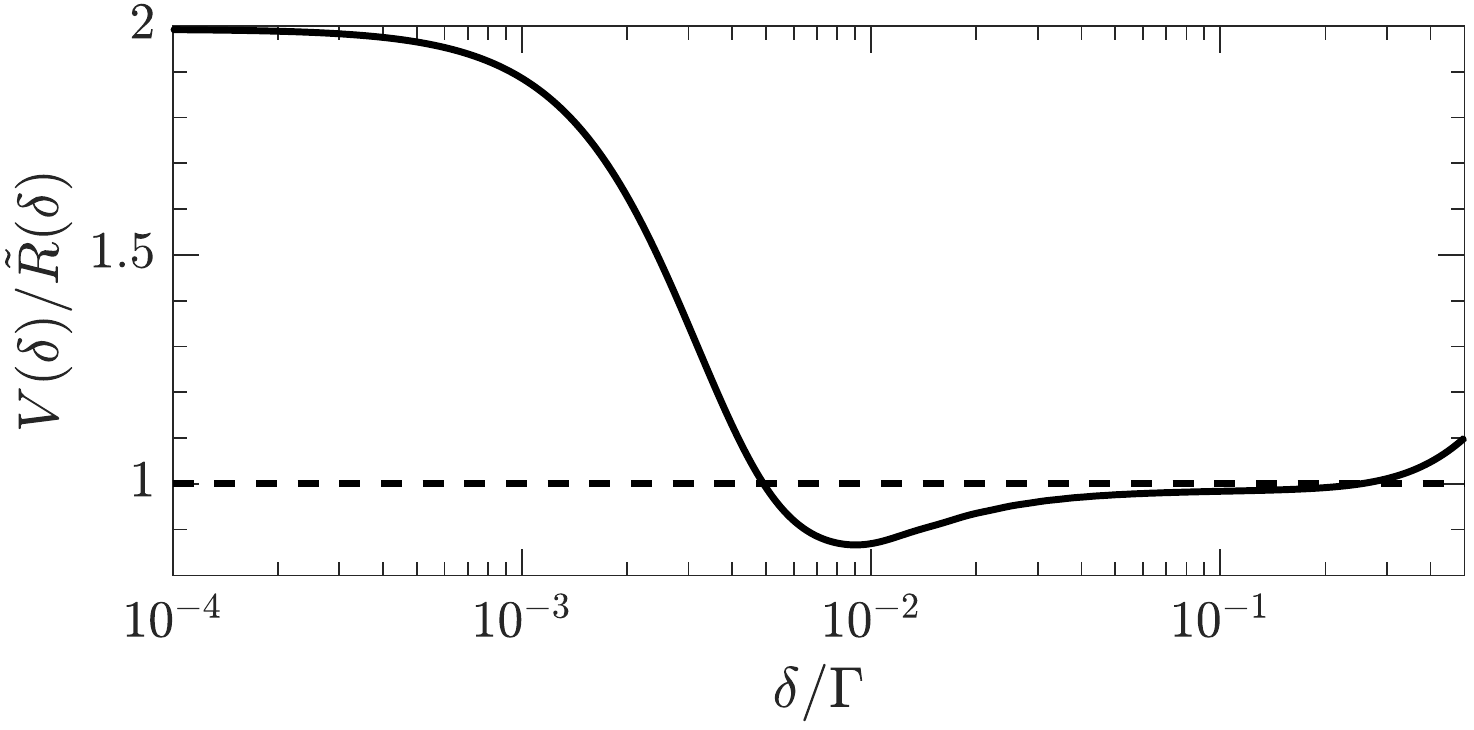}
}

\subfloat[\label{fig:compareB}]{
\includegraphics[trim=0 0 0 0,width=0.9\columnwidth]{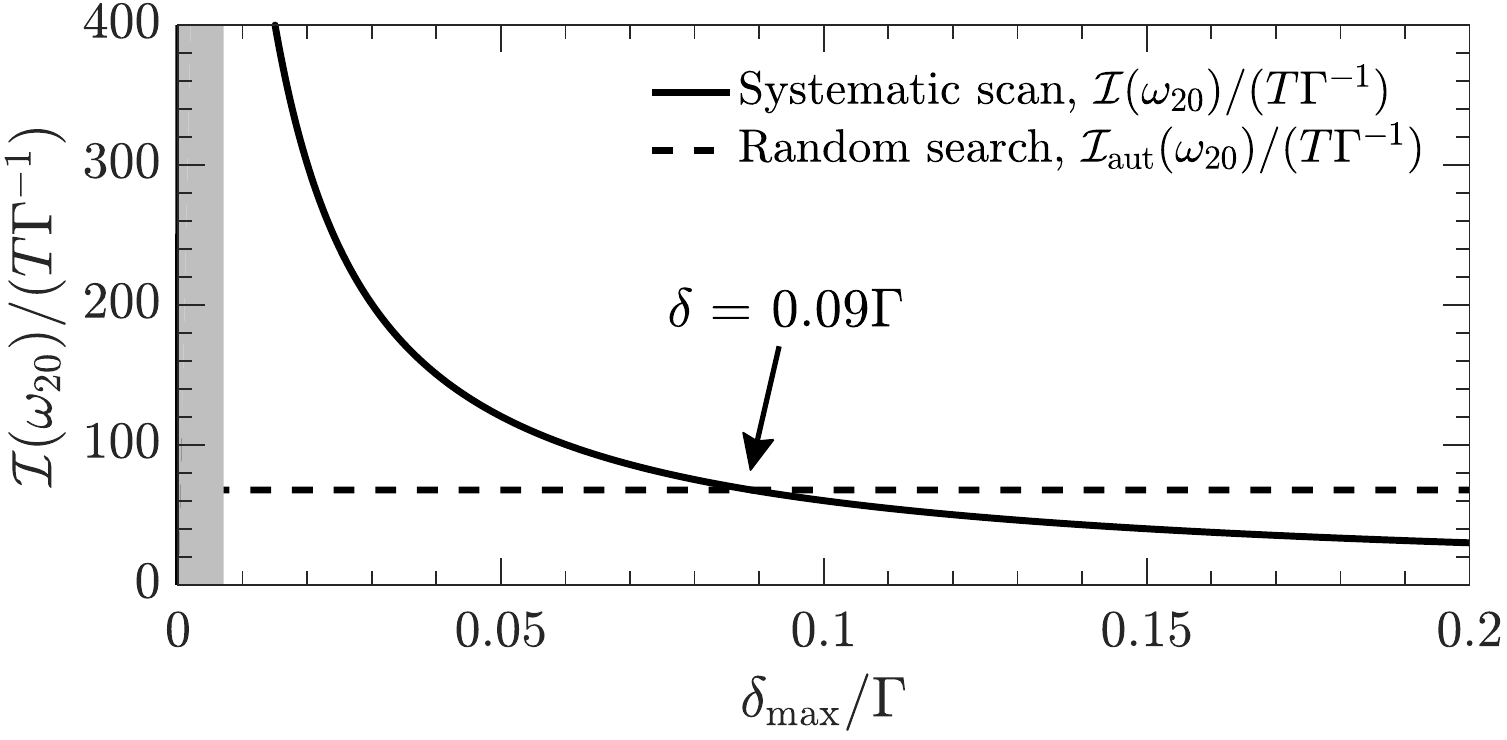}
}
\caption{(a) Dependence on the detuning $\delta$ of the photocount variance per time $V(\delta)$ divided by the rate $\tilde{R}(\delta)$. The dashed line
marks the Poissonian case where $V(\delta)=\tilde{R}(\delta)$.
(b) Information measures for estimating $\omega_{20}$ in the $\Lambda$-system by a systematic scan and the random search protocol, respectively. In both cases the search is restricted to an interval $[-\dM,\dM]$ around the dark resonance.
The shaded area is the region where $\dM<\dL$ and our statistical model of the recycling process requires modifications.
Results are shown for $\Omega = 0.1\Gamma/\sqrt{2}$.
}
\label{fig:compare}
\end{figure}

From \eqref{eq:vk}, we notice that the second term in \eqref{eq:discreteFisher} does not scale with $T$ and is hence negligible at large times.
Taking the limit of a continuum of frequencies, $N\rightarrow \infty$, we transform the sum in \eqref{eq:discreteFisher} to an integral and obtain our final expression for the Fisher information of estimating a parameter $\theta$ by systematically scanning a laser frequency across a resonance,
\begin{align}\label{eq:Fisher}
\mathcal{I}(\theta)= \frac{T}{2\dM}\int_{-\dM}^{\dM} d\delta \, \frac{1}{V(\delta)} \left[\pdiff{\tilde{R}(\delta)}{\theta}\right]^2,
\end{align}
where
\begin{align}
V(\delta) = \tilde{R}(\delta)+2\sum_i\int_{0}^{\infty}d\tau\, \tilde{G}_i^{(2)}(\tau)
\end{align}
is the frequency dependent photocount variance per time.
The Fisher information \eqref{eq:Fisher} reveals via the Crámer Rao bound an uncertainty
$\sigma(\theta)= \left[\mathcal{I}(\theta)\right]^{-1/2}$, scaling as $1/\sqrt{T}$ with time.

In \fref{fig:compareA} we show  $V(\delta)/\tilde{R}(\delta)$ as a function of the detuning $\delta$ for the $\Lambda$-system considered in the main text. Notice how the photo current exhibits photon bunching and super-Poissonian counting statistics close to the dark resonance, while it is sub-Possonian for intermediate values and again super-Poissonian away from the resonance.

To compare with the autonomous search protocol presented in the main text, we show in \fref{fig:compareB} the Fisher information \eqref{eq:Fisher} for estimating $\omega_{20}$ by a systematic scan along with the
equivalent information measure $I_\mathrm{aut}(\omega_{20}) = \left(\dTh/0.82\right)^2$ of our random search for different widths of the search interval as determined by $\dM$.
The comparison shows that
for the parameters used in \fref{fig:trajectory}, our random search proves superior to the frequency scan if we search an interval $[-\dM,\dM]$ with $\dM>0.09\Gamma$ , i.e. as long as the resonance is not a-priori known to very high precision.

\section{Outlook}\label{sec:Outlook}
While we presented the scheme for a driven $\Lambda$-system and restricted our attention to a rate $R(\delta)$ with a quadratic dip around $\delta=0$ and a flat plateau for large $\delta\simeq \dM$, the arguments are general, and the statistical methods apply equally well to other systems.
For example, different forms of $R(\delta)\propto \delta^{\alpha}$ for $\delta\simeq 0$ will lead to different values of $\mu=1/\alpha$ which, in turn, imply different scaling with time of the sensitivity as quantified by $\dTh\propto T^{-\mu}$.
For instance, a variant of the presented scheme may apply Raman pulses rather than continuous illumination. It can be shown that e.g. a sequence of Blackman pulses \cite{1455106} yields an excitation probability characterized by $\mu=1/4$, while square pulses lead to $\mu=1/2$ \cite{PhysRevLett.75.4575}.
Although these examples do not yield a faster convergence of the random walk in frequency space towards the atomic resonance frequency than the example studied here, they illustrate the usefulness of the general formalism. 
This formalism will allow better than $1/\sqrt{T}$ scaling of the error in estimating a general unknown parameter $\theta$, if a process is found 
for which the rate depends on $\theta$ as $R(\theta)\propto \theta^{\alpha}$ with $0< \alpha < 2$.

If $\omega_{20}$ is only known to a precision of $\lesssim \Gamma$, $\dM$ must be chosen bigger than $\dLor$. In this case, the rate decreases as $1/\delta^2$ in the recycling region leading to recycling times of order $\sim\delta^2$, and there is a risk that trajectories will be trapped far away from the resonance.
The return times are then also described by anomalous statistics, and $P_r(\tau_r)$ is of the form \eqref{eq:power} with $\mu_r<1$. The actual value of $\mu_r$ depends on the detailed frequency-shifting protocol.
If $\delta$ is restricted to jump to a vicinity of the current value, one finds $\mu_r=1/4$ \cite{Levy} and  $T^{(REC)}_N \propto N^4$. Our scheme then fails asymptotically as $f_E(T)\rightarrow 0$ for large $T$. If, instead, the laser frequency is shifted uniformly on the search interval, the exact zero of $R(\delta)$ at $\delta=0$ dominates the asymptotic zero as $|\delta|\rightarrow\infty$, and the trajectories will converge (albeit more slowly) to the PDS.

In this work, we have proposed to locate the absorption zero of a dark resonance by a random frequency search protocol. Due to the non-ergodic behavior of the system, methods from L\'{e}vy statistics were employed to assess the asymptotic spectroscopic sensitivity of the scheme. For the example of a driven $\Lambda$-type system, our method compares favourably with the Cram\'{e}r-Rao bound of a conventional frequency scan. Metrology protocols have been proposed, that feature similar feedback and adaptive elements, and which show convergence faster than $1/\sqrt{T}$ or $1/\sqrt{N}$, where $N$ quantifies the amount of physical resources; see e.g. \cite{PhysRevLett.96.010401,higgins2007entanglement,PhysRevA.94.042121,PhysRevA.94.023840}. Since adaptive schemes may generally induce non-ergodic dynamics, we believe that elements of our theoretical analysis will be relevant in the characterization of a number of such protocols where standard statistical analyses are inadequate.
\begin{acknowledgments}
The authors acknowledge financial support from the Villum Foundation.
A. H. K. further acknowledges support from the Danish Ministry  of  Higher  Education  and  Science.
\end{acknowledgments}

\section*{appendix}\label{sec:appendix}
\appendix

\section{Effective emission rate for a laser driven $\Lambda$-system}
\label{sec:R}

The laser driven $\Lambda$-system in \fref{fig:setup} of the main text is described by the Hamiltonian
\begin{align}
\hat{H} =  \delta \ket{0}\bra{0}
+\frac{\Omega}{2}\left(\ket{2}\bra{0}+\ket{0}\bra{2}\right)
+\frac{\Omega}{2}\left(\ket{2}\bra{1}+\ket{1}\bra{2}\right),
\end{align}
with laser atom detuning $\delta$ and Rabi frequency $\Omega$.

The evolution of the density matrix $\rho$ of the unobserved system is given by the master equation $d\rho/dt = L [\rho]$,
where the Liouvillian superoperator is defined by ($\hbar=1$)
\begin{align}\label{eq:liovillian}
L[\rho] = -i[\hat{H},\rho]+\sum_i\left( \hat{c}_i\rho \hat{c}_i^\dagger -\frac{1}{2} \left\{\hat{c}_i^\dagger \hat{c}_i,\rho \right\}\right).
\end{align}
Here the excited state spontaneous decay with rate $\Gamma$ is represented by the relaxation operators $\hat{c}_{0} = \sqrt{\Gamma/2}\ket{0}\bra{2}$ and $\hat{c}_{1} = \sqrt{\Gamma/2}\ket{1}\bra{2}$.
The unobserved system relaxes to a steady state $\rst$ with $L[\rst]=0$ from which follows the average properties of the emitted radiation. In particular, the average fluorescence rate is given by $\tilde{R}(\delta) =\sum_i \Tr{\hat{c}_i^\dagger\hat{c}_i \rst}$ where $i = 0,1$, and the quantum regression theorem yields Glauber's correlation function \cite{PhysRev.130.2529} for two photo emissions in channel $i$ separated by a time $\tau$,
\begin{align}
G_i^{(2)}(\tau) = \Tr{\hat{c}_i^\dagger\hat{c}_i\e{L\tau}\left[\hat{c}_i\rst\hat{c}_i^\dagger\right]}.
\end{align}

\begin{figure}
\includegraphics[trim=0 0 0 0,width=0.9\columnwidth]{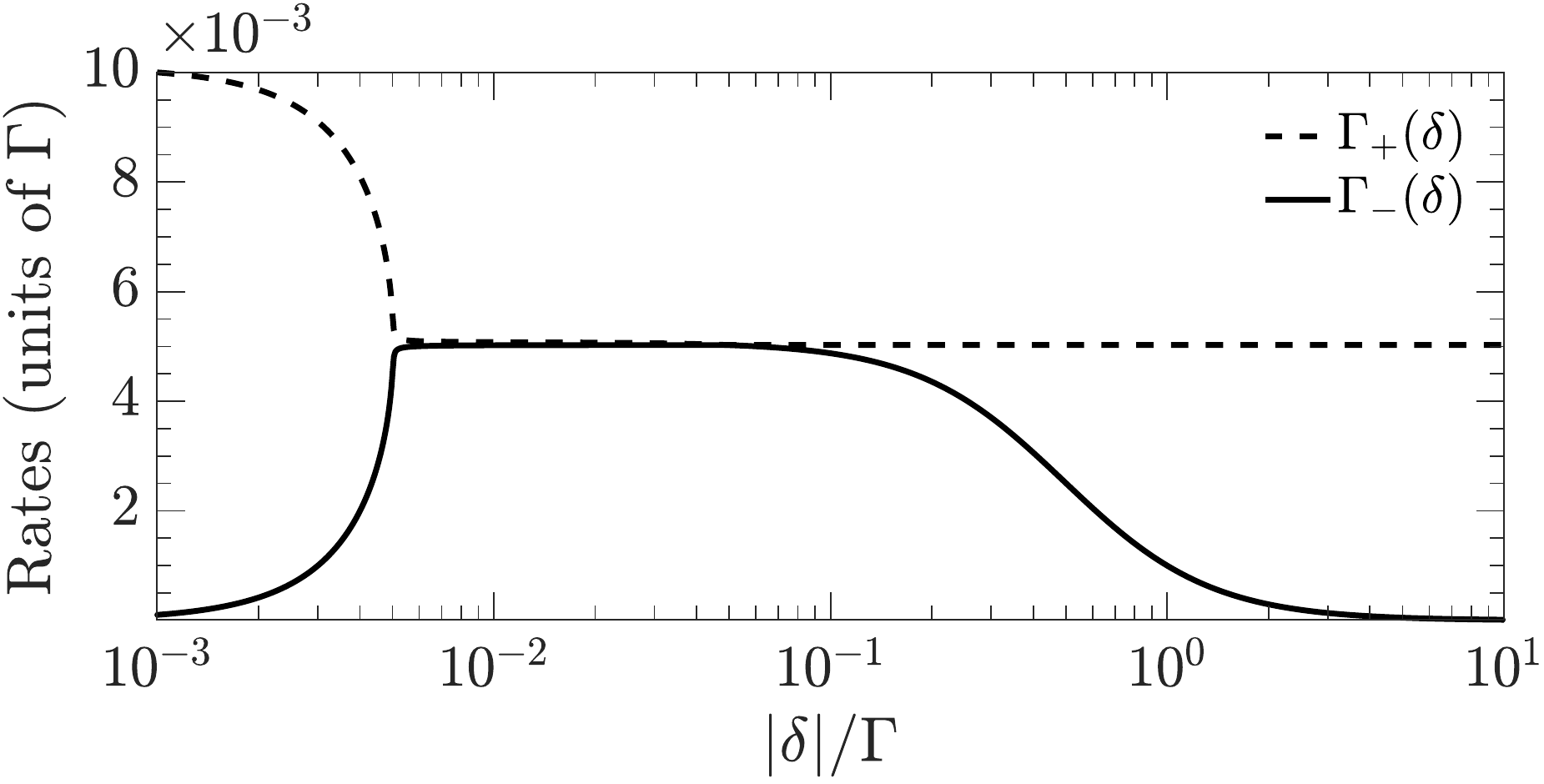}
\caption{Frequency dependent effective emission rates from the bright $\ket{\psi_+}$ and dark $\ket{\psi_-}$ state superpositions of the two ground states of a $\Lambda$-type system. The rates are even functions of $\delta$ and results are shown for $\Omega = 0.1\Gamma/\sqrt{2}$.
}
\label{fig:EigenvaluesHeff}
\end{figure}

The master equation can be unravelled into stochastic evolution corresponding to the random measurement back action on the system due to detection of the emitted radiation by photon detectors. Whenever a photon is detected, the system jumps to the corresponding ground state, $|\psi\rangle \rightarrow \hat{c}_i |\psi\rangle/\sqrt{\langle\psi|\hat{c}^\dagger_i\hat{c}_i|\psi\rangle}$,  while between photo detections the evolution of the (unnormalized) state $\ket{\tilde{\psi}}$ is governed by an effective Hamiltonian,
\begin{align}
\hat{H}_{\text{eff}} = \hat{H}-\frac{i}{2}\sum_i \hat{c}_i^\dagger\hat{c}_i,
\end{align}
where the imaginary term represents the decay of the excited state.
The eigenstates of $\hat{H}_{\text{eff}}$, $\ket{\psi_j}$ with eigenvalues $\lambda_j$ represent decaying modes with decay rates $\Gamma_j= -2\mathrm{Im}(\lambda_j)$.
For weak driving, the largest of these rates is almost equal to $\Gamma$ and the corresponding eigenstate is close to the bare atomic excited state, i.e., it has negligible statistical weight $w_j^{(n)}=|\langle n\ket{\psi_j}|^2$ on the atomic ground states ($n=0,1$).
The two smallest rates $\Gamma_-$ and $\Gamma_+$, on the other hand, are associated with the ground states, and hence they constitute the effective fluorescence rate right after a detector click. Their dependence on the detuning $\delta$ is shown in \fref{fig:EigenvaluesHeff}.
Close to resonance $\Gamma_-(\delta)$ tends quadratically to zero, while $\Gamma_+(\delta)$ increases equivalently.
This is because $\Gamma_-(\delta\simeq 0) \simeq 0$ corresponds to the dark state superposition $\ket{\psi_{-}}=\left(\ket{0}-\ket{1}\right)/\sqrt{2}$ while $\Gamma_+(\delta\simeq 0)\simeq 2\Omega^2/\Gamma$ is the rate of excitation and emission from the bright state linear combination $\ket{\psi_{+}}=\left(\ket{0}+\ket{1}\right)/\sqrt{2}$.

Upon photo detection, the atom may, with probability $w_+^{(0,1)}\simeq 1/2$, continue to fluoresce at a rate $\Gamma_+(\delta)$, and hence quickly remit, but it may also, with a probability $w_-^{(0,1)}\simeq 1/2$, continue to fluoresce at a rate $\Gamma_-(\delta)$ corresponding to the pseudo dark state.
The frequency dependent emission rate leading to non-ergodic dynamics close to resonance is thus given by $R(\delta) = \Gamma_-(\delta)$.

Our Lévy statistical analysis relies on the overall rather than the detailed shape of the emission rate from the ground states. In this spirit we note that the fluorescence rate $\Gamma_-(\delta)$ as a function of the detuning is characterized by a dip with quadratic variation around $\delta=0$ due to the dark state, a plateau where the rate is constant, and a tail where the decay follows a Lorentzian line shape due to off-resonant scattering. We hence approximate the rate by
 \begin{equation} \label{eq:R}
 R(\delta)=\left\{
             \begin{array}{ll}
               \tau_0^{-1}(\delta/\dL)^2, &\quad |\delta| < \dL \\
               \tau_0^{-1}, &\quad \dL < |\delta| < \dLor \\
               \tau_0^{-1}(\dLor/\delta)^2,  &\quad \dLor < |\delta|.
             \end{array}
           \right.
\end{equation}
where the characteristic parameters are identified by matching
the plateau to the maximum of $\Gamma^{(-)}(\delta)$, and requiring that $R(\delta)$ represents the exact form in the limits $\delta\simeq 0$ and $\delta\gg 0$.
Though not a necessity for our analysis, we assume for simplicity that the coupling is weak ($\Omega\ll \Gamma$). We then find by applying second order perturbation theory that
\begin{align}
\tau_0 &= \frac{\Gamma}{\Omega^2} \nonumber
\\
\dL &= \sqrt{2}\frac{\Omega^2}{\Gamma}
\\
\dLor &= \frac{\Gamma}{2}. \nonumber
\end{align}
The approximation \eqref{eq:R} is compared to the exact rate in Figure 2 of the main text.

\section{Broad distributions and Lévy statistics}\label{sec:ABroad}
In this section we give a brief introduction to 'broad distributions' decaying slowly at large deviations.
We will focus on the typical cases of power-law decays. Let $\tau$ be a positive random variable distributed for large values according to
\begin{equation} \label{eq:powerSup}
P(\tau) \underset{\text{large } \tau}{\simeq} \frac{\mu \tau_b^{\mu}}{\tau^{1+\mu}},
\end{equation}
where the exponent $\mu$ determines the decay of the tail towards zero.
Normalizability requires $\mu>0$ and all moments $\braket{\tau^n }$ for which $n\geq \mu$ diverge.

The central limit theorem (CLT) concerns the asymptotic behavior of the sum $T_N$ of $N$ independent realizations of the probability distribution $P(\tau)$,
\begin{align}\label{eq:TN}
T_N = \sum_{i=1}^N \tau_i,
\end{align}
independent of the detailed shape of $P(\tau)$.

For $\mu>2$ both the mean $\braket{\tau}$ and variance $\sigma^2$ of $\tau$ are finite, and defining a random variable $\epsilon$ such that
\begin{align}
T_N = \braket{\tau}N+\epsilon\sigma \sqrt{N},
\end{align}
the normal CLT ensures that for large $N$, $\epsilon$ is a Gaussian random variable with zero mean and unit variance. I.e. $T_N$ is normally distributed and $T_N \rightarrow \braket{\tau}N$ for large $N$.

For $\mu<2$ the variance of $\tau$ is formally infinite, and the normal CLT does not apply. Instead a generalized CLT has been proven by Lévy and Gnedenko \cite{PaulLevy,bouchaud1990anomalous}.
If $1<\mu<2$ the mean value of $\tau$ is finite, and by defining the Lévy increment $\xi$ such that
\begin{align}\label{eq:TL}
T_N = \braket{\tau}N+\xi\tau_b N^{1/\mu},
\end{align}
the generalized CLT states that $\xi$ is a random variable of order $1$ distributed according to the completely asymmetric Lévy distribution $L_\mu(\xi)$ which only depends on the value of $\mu$. Notice, however, that we still have $T_N \rightarrow \braket{\tau}N$ for large $N$.

The most interesting case in the present work is $\mu<1$, where even the mean of $\tau$ is undefined.
Then \eqref{eq:TL} becomes
\begin{align}
T_N = \xi\tau_b N^{1/\mu},
\end{align}
where $\xi$ is distributed as above, and we note that the sum \eqref{eq:TN} no longer scales proportionally with the number of terms, but rather is dominated by a few single terms.

\section{Recycling time distribution}\label{sec:Arecycling}
Here we address the temporal dynamics and derive the probability distribution $P_r(\tau_r)$ of the recycling time intervals $\tau_r$.

We introduce first the probability $P_{1}(k)$ that the detuning returns to the PDS (defined in the main text) for the first time at exactly $k$ photon detection events after leaving the PDS. Notice that $P_{1}(k)$ relates to the number of jumps and \emph{not} to the duration $\tau_r$ of the time spent outside the PDS.
The probability $\PT(n)$ that the system occupies the PDS after the $n^{th}$ detection event can be written as
a sum over probabilities of already being trapped after $n'$ events with probability $\PT(n')$, leaving the PDS at $n'+1$ (which occurs with unit probability since $\dT\ll \dM$) and returning after an additional $n-n'$ steps with a probability $P_1(n-n')$,
\begin{align}\label{eq:PT}
\PT(n) = P_1(n)+\sum_{n'=0}^n \PT(n')P_1(n-n'),
\end{align}
where the first term accounts for a first return at $n$ without any prior returns.
We assume an initial detuning in the recycling region, and we have extended the summation limits to $n'=0$ and $n'=n$ which is justified since $\PT(0)=0$ and $P_1(0) = 0$.

The sum constitutes a convolution product, and we introduce the discrete Laplace transform (moment-generating function),
\begin{align}\label{eq:Laplace}
\mathcal{L}_d P(s) = \sum_{n=0}^\infty \e{-sn} P(n)
\end{align}
realizing the relation between $P_1(n)$ and $P_{\text{trap}}(n)$,
\begin{align}\label{eq:P1PT}
\mathcal{L}_d P_1(s)=\frac{\mathcal{L}_d\PT(s)}{1+\mathcal{L}_d \PT(s)}.
\end{align}
This result is independent of any specific frequency-shifting protocol.

The main text investigates the case where after each detection event the detuning explores the interval $\delta\in[-\dM,\dM]$ in a uniform manner. In such settings, $\PT(n)$ has a constant value
\begin{align}
P_{\text{trap}}(n) = \frac{\dT}{\dM},
\end{align}
and Eq.~(\ref{eq:P1PT}) yields
\begin{align}
\mathcal{L}_d P_1(s)= 1-\frac{\dM}{\dT}s.
\end{align}
Since $\mathcal{L}_d P_1(s)$ is a moment-generating function this implies that the average number of steps before the first return is finite and given by
\begin{align}
\braket{n} = \frac{\dM}{\dT}.
\end{align}

The temporal duration of each step depends on the emission rate in the recycling region. In the main text we focus on the case $\dL\ll\dM<\dLor$, where the recycling region is characterized by a frequency independent rate, $R(\delta)=1/\tau_0$, and the average time $\tau_0$ between two jumps is finite. The average first return time is then simply
\begin{align}
\braket{\tau_r} = \braket{n}\tau_0.
\end{align}
The finite mean value implies that the recycling times $\tau_r$ follow normal statistics. In fact, it can be shown that the tail of $P_r(\tau_r)$ follows an exponential law \cite{Wei94}.

If the frequency-shifting is performed as an unconfined standard random walk
\eqref{eq:P1PT} still applies and leads to a first return distribution with a power law tail
\begin{align}
P_1(n)\underset{\text{large } n}{\simeq}\frac{1}{2\sqrt{2\pi}}\frac{\Delta \delta}{\dT}\frac{1}{n^{3/2}},
\end{align}
with $\Delta \delta$ the average step size \cite{Levy}. In this case $\braket{n}$ diverges.
The corresponding statistical behavior of the recycling times $\tau_r$
is dominated by trapping in effective dark states at high $\delta$ where, by \eqref{eq:R}, $R(\delta)\propto 1/\delta^2$.
One finds \cite{Levy} that $P_r(\tau_r)$ then follows \eqref{eq:powerSup} with  $\mu_r = 1/4$ and $\tau_{r,b}= \tau_0 (\Delta \delta)^6/(\dT^4\dLor^2)$, and that the recycling process is dominated by very long time intervals.

\section{Proportion of trapped trajectories}\label{sec:AtrapProportion}
Here we derive the proportion of trajectories that will asymptotically for long times be trapped in the PDS with $|\delta|<\dT$.
Due to the non-ergodic dynamics, the time average, unlike the ensemble average results, retains a stochastic contribution even in the long time limit.

The alternation between trapping and recycling periods defines a renewal process \cite{cinlar1969markov}, and we introduce first the probability density functions $S_R(t)$ of returning to the PDS region at time $t$ independent of the number of previous return points and $S_D(t)$ for departing at time $t$ independent of previous departure points.
I.e. $S_R(t)dt$ ($S_D(t)dt$) is the probability of entering (departing) the PDS region in $[t,t+dt]$.
The densities can be expressed in terms of each each other and the trapping and recycling time distributions. For an initially un-trapped trajectory, we have
\begin{align}\label{eq:SR}
S_R(t) = P_r(t)+\int_0^t d\tp\, S_D(\tp)P_r(t-\tp),
\end{align}
where the first term accounts for the probability of being trapped exactly at $t$ and the second the case of escaping at $\tp\in[0,t]$ and returning at $t$.
Similarly
\begin{align}\label{eq:SD}
S_D(t) = \int_0^t d\tp \,S_R(\tp)P_t(t-\tp).
\end{align}
The integrals in the expressions (\ref{eq:SR},\ref{eq:SD}) form convolution products, so performing Laplace transforms, $\mathcal{L}g(s) = \int_0^{\infty} dt\, g(t)\e{-st}$, and eliminating $\mathcal{L} S_D(s)$, we find
\begin{align}\label{eq:LSR}
\L S_R(s) = \frac{\L P_r(s)}{1-\L P(s)\L P_r(s)}.
\end{align}

The ensemble average trapped proportion at time $T$ can be written as an integral over time $\tp$ of the probability that the system entered the trap at time $\tp$ multiplied by the probability $\psi(T-\tp)$ that the system remained in the trap until times later than $T$,
\begin{align}\label{eq:feSupp}
f_E = \int_0^T d\tp \,S_R(\tp) \psi(T-\tp).
\end{align}
Note that $\psi(T-\tp)$  is itself an integral over the distribution $P_t(T-t')$ of trapping times,
\begin{align*}
\psi(\tau) = \int_\tau^\infty d\tau^\prime \, P_t(\tau^\prime).
\end{align*}
The Laplace transform of the convolution \eqref{eq:feSupp} is
\begin{align}
\mathcal{L}f_E(s) = \L S_R(s)\L\psi(s),
\end{align}
with $\L\psi(s)=(1-\L P(s))/s$. Inserting \eqref{eq:LSR} we thus reach our final expression for the Laplace transform of the trapped proportion,
\begin{align}\label{eq:Lfe}
\L f_E(s) = \frac{\L P_r(s)}{1-\L P_t(s)\L P_r(s)}\frac{1-\L P_t(s)}{s},
\end{align}
revealing
\begin{align}
f_E(T) =\int_0^T dt\,\left[S_R(t)-S_E(t)\right],
\end{align}
which is very sensible.

With $\mathcal{L}f_E(s)$ expressed in terms of the trapping and recycling time distributions we may apply our statistical model.
A small $s$ expansion (high $\tau_t$) of the Laplace transform of $P_t(\tau_t)$ as given in Eq. (2) of the main text yields to first order \cite{Levy}
\begin{align} \label{eq:LP}
\L P_t(s) \simeq 1- \Gamma(1-\mu)(s\tau_b)^\mu,
\end{align}
where $\Gamma(x)$ is the Gamma-function.
For the recycling distribution we focus on the case $\dM<\dLor$, where the mean recycling time is finite so that
\begin{align}
\L P_r(s) = 1-s\braket{\tau_r}
\end{align}
for small $s$.
Then by \eqref{eq:Lfe}
\begin{align}
\L f_E(s) = \frac{1}{s}-\frac{\braket{\tau_r}}{\Gamma(1-\mu)(s\tau_b)^\mu},
\end{align}
and one can finally show that asymptotically as $T\rightarrow \infty$ the inverse transform gives
\begin{align}
f_E(T) \simeq 1- \frac{\sin(\pi\mu)}{\pi}\frac{\braket{\tau_r}}{ \tau_b^\mu T^{1-\mu}}.
\end{align}
For a discussion of cases in which $\dM>\dLor$, the reader is referred to \cite{Levy}.

\section{Asymptotic frequency distribution}\label{sec:AFrequencyDistribution}
The asymptotic proportion of trajectories with $|\delta|<\dT$ is given by $f_E(T)$. The asymptotic distribution $\mathcal{P}(\delta,T)$ of this proportion is found by integrating the probability of entering the trap at a time $\tp$ with a given $\delta$ and not leaving before the final time $T$,
\begin{align}
\mathcal{P}(\delta,T) = \rho(\delta)\int_0^T d\tp \, S_R(\tp)\phi(T-\tp|\delta),
\end{align}
where $\rho(\delta)=1/2 \dT$ is normalized, so $f_E(T)=\int_{-\dT}^{\dT} d\delta \,\mathcal{P}(\delta,T)$, and we define the probability to leave the trap after a time $\tau$ conditioned on the value of $\delta$,
\begin{align}
\phi(\tau|\delta) = \int_\tau^\infty d\tau^\prime \, P_t(\tau^\prime|\delta).
\end{align}
As it turns out, the time-dependent distribution of frequencies $\delta\leq \dT$ \textit{within} the trap is self-similar for different times and can in general be factorized as
\begin{align}
\mathcal{P}(\delta,T)= h(T)G(q).
\end{align}
We restrict our attention to the case $\dM<\dLor$ with infinite average trapping time and finite recycling times, and we refer to \cite{Levy} for derivations when the recycling is also non-ergodic.
From \eqref{eq:LSR} and \eqref{eq:LP} it follows that the small $s$ expansion of the Laplace transform of the renewal density function is $\mathcal{L}S_R(s) = (s\tau_b)^{-\mu}/\Gamma(1-\mu)$, so that for large times
$S_R(t) \simeq \sin(\pi\mu)\tau_b^{-\mu}t^{\mu-1}/\pi$.
One finds then the height of the distribution,
\begin{align}
h(T)=\left(\frac{\tau_{PDS}}{\tau_b}\right)^\mu\frac{\sin(\pi\mu)}{\pi\mu \dTh}.
\end{align}
The form factor is defined as a function of $q=\delta/\dTh$ as
\begin{align}\label{eq:G}
G(q)=\mu\int_0^1 du\, u^{\mu-1}\e{-(1-\mu)q^{1/\mu}},
\end{align}
which for $\mu=1/2$ can be expressed as $G(q)=D(q)/q$, where $D(q)$ is the Dawson function.
The tails of $G(q)$ are like a Lorentzian $\sim 1/2q^2$ and the area is $\pi^{3/2}/2$.
\begin{figure}[h]
\includegraphics[trim=0 0 0 0,width=0.95 \columnwidth]{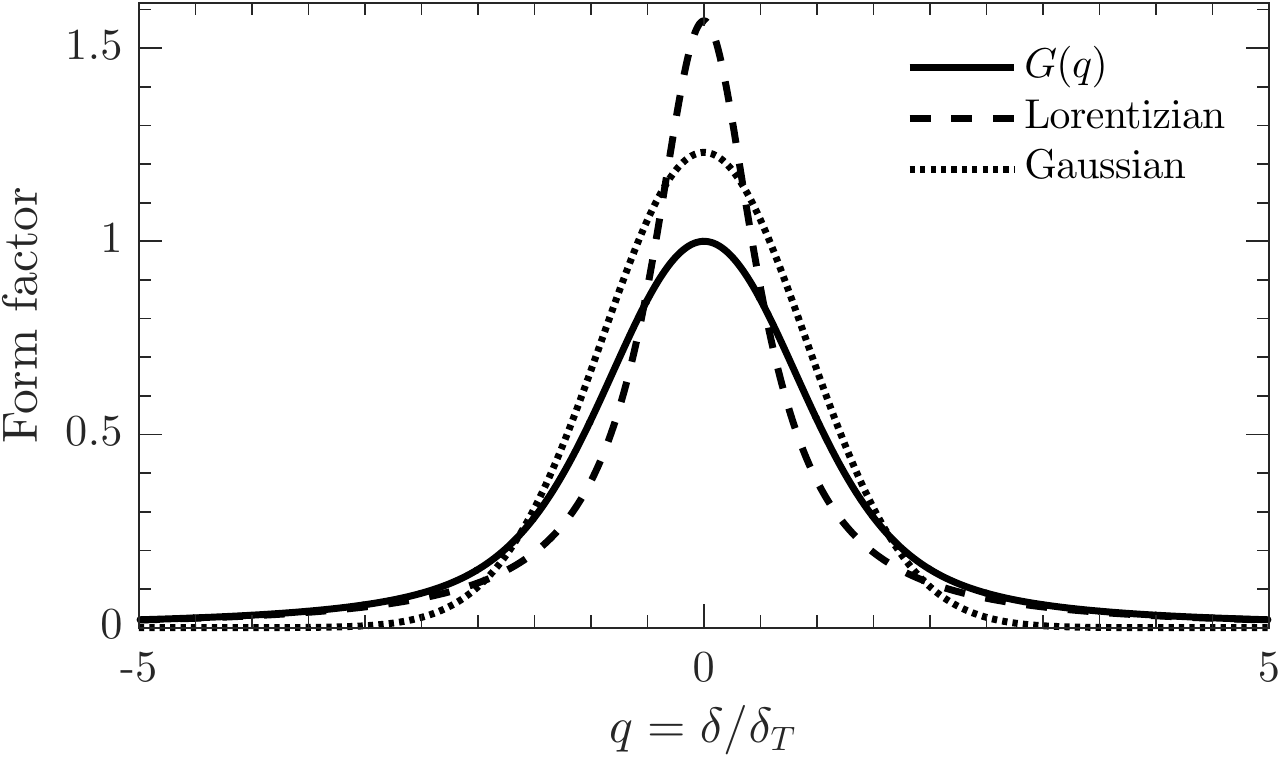}
\caption{The form factor $G(q)$ \eqref{eq:G} of $\mathcal{P}(\delta,T)$ is compared to a Lorentzian with the same tails ($\propto 1/2q^2$), and a Gaussian with the same FWHM ($2.13$).
All distributions are normalized to an area $\pi^{3/2}/2$.}
\label{fig:G}
\end{figure}
$G(q)$ is compared to a Lorentzian with the same tails and a Gaussian with the same FWHM and normalization in \fref{fig:G}.
Notice that the distribution is not as narrow as the Lorentzian close to the central frequencies.

The resulting properties of $\mathcal{P}(\delta,T)$ are discussed in the main text.

\bibliography{/home/alexander/Dropbox/PhD/Bibliography/master}{}

\end{document}